# The pressure field beneath intense surface water wave groups

A. Slunyaev[1,2], E. Pelinovsky[1,2,3] and H.-C. Hsu[4]

1) Institute of Applied Physics, 46 Ulyanova Street, Nizhny Novgorod 603950, Russia
2) Nizhny Novgorod State Technical University, 24 Minina Street, N. Novgorod 603950, Russia
3) National Research University-Higher School of Economics, 25 B. Pechorskaya Street, Nizhny Novgorod 603950 Russia
4) Tainan Hydraulics Laboratory, National Cheng Kung University, Tainan, 701 Taiwan, Republic of China

Slunyaev@appl.sci-nnov.ru

**Abstract**
A weakly-nonlinear potential theory is developed for the description of deep penetrating pressure fields caused by single and colliding wave groups of collinear waves due to the second-order nonlinear interactions. The result is applied to the representative case of groups with the sech-shape of envelope solitons in deep water. When solitary groups experience a head-on collision, the induced due to nonlinearity dynamic pressure may have magnitude comparable with the magnitude of the linear solution. It attenuates with depth with characteristic length of the group, which may greatly exceed the individual wave length. In general the picture of the dynamic pressure beneath intense wave groups looks complicated. The qualitative difference in the structure of the induced pressure field for unidirectional and opposite wave trains is emphasized.

## 1. Introduction

The sea surface motions are accompanied by complicated movements of fluid particles within the entire water column. The underwater fluid dynamics may have significant effect on submerged constructions, formation of bottom topography peculiarities, sediment transport and so on. Obviously, extreme surface waves may cause stronger underwater effects; enormous wave force records at coastal and moored structures are sometimes reported (e.g. Haver, 2005) and are often related to so-called rogue wave events (see Kharif et al, 2009). Thus, the problem of estimation of underwater extreme pressures and the associated impact on structures has clear practical importance.

On the other hand, subsurface and bottom pressure sensors are commonly used for the registration of surface waves in relatively low-frequency range, from tidal to wind waves. This type of wave recorders is very appropriate as they do not deface the scenery in coastal areas and are difficult to be accessed by unauthorized people (vandal-proof). We may speculate that arrays of submerged probes also could be used for safe and inconspicuous recording of wind waves.

The inverse problem of an accurate recovery of characteristics of wind waves from the data of subsurface/bottom pressure measurements is vital. The pressure under very long waves such as tides and tsunamis may be assumed hydrostatic; this approximation is sufficient for an accurate reconstruction of the surface waves from the bottom pressure sensor data, what enables creating of tsunami early warning networks. However, wind waves are often not sufficiently long even in the coastal area. Then non-hydrostatic corrections to the pressure contribute significantly and the relation between the surface displacement and the pressure at the location of registration becomes complicated. As the inverse problem is highly attractive, it has motivated a number of studies. The linear theory based on the spectral methods was developed in [Cavaleri, 1980; Wang et al, 1986; Bishop & Donelan, 1987; Kuo & Chiu, 1994; Baquerizo & Losada, 1995; Zaslavsky & Krasitsky, 2001; Tsai et al, 2005; Huang & Tsai, 2008], though the importance of nonlinear effects was demonstrated shortly after. Most of the studies concern shallow water conditions; it is reasonable that representative solutions such as uniform waves, shallow water solitons and their interactions are thoroughly studied (Escher & Schlurmann, 2008; Constantin et al. 2010, 2011; Constantin, 2012; Oliveras et al, 2012; Clamond & Constantin, 2013; Touboul & Pelinovsky, 2014, 2016; Pelinovsky et al, 2015), and model theories are being validated against the direct solution of hydrodynamic equations.

The solutions of the Laplace equation for potential planar surface waves inevitable possess the feature that each surface wave harmonic with a given horizontal scale attenuates with depth with exactly the same vertical scale. Thus, short waves have no effect on sufficiently deep water layers. This picture gets broken due to the nonlinear wave interactions which may lead to the generation of longer wave components, which penetrate much deeper into the water bulk. This mechanism was suggested by Longuet-Higgins (1950) to explain the seismic noise generated by wind waves, which would not produce noticeable effect on the sea bottom over relatively deep areas in the linear regime. Longuet-Higgins (1950) suggested a simple solution for two counter propagating uniform waves (i.e., standing waves), which due to the nonlinear interaction produce a harmonic with the frequency $\omega_0 + \omega_0 = 2\omega_0$ and the wavenumber $k_0 - k_0 = 0$. As a result, the second-order pressure does not attenuate at large depth at all. A more general treatment of the problem when the waves obey some spectrum was given by Hasselmann (1963). The approach by Longuet-Higgins (1950) seems to prevail in the literature probably due to its simplicity. A review of the nonlinear sea wave interaction theory in the context of a seismic wave generation may be found in monograph by Kibblewhite & Wu (1996). The effect of sea surface waves on a seismic noise is usually considered in the statistical sense, when the generated seismo-acoustic spectra are discussed.

In this paper we consider the underwater pressure fields beneath isolated intense wave groups. The groups own a new characteristic length scale, which is typically $O(10)$ times larger than the wave length. Due to the nonlinear wave interactions the modulated waves generate large-scale perturbations which decay with depth with the typical scale of the group length (Dysthe, 1979; Yezersky et al, 1981; McIntyre, 1981). In that way, a given water basin in terms of group lengths is not as deep as in terms of individual wave lengths.

The group structure of sea waves is specified not solely by the given wave spectrum, but also by the dynamical effects of the modulations. In particular, in deep water regions intense waves tend to split into the wave groups characterized by an approximate balance of dispersive and nonlinear effects (i.e., envelope solitons, nonlinear solitary groups, Benjamin & Feir, 1967; Zakharov, 1968). The experimental observations of intense solitary groups are reported in Slunyaev et al (2013, 2017). The balance between the nonlinearity and dispersion allows the groups to approximately preserve the shape for long time. Therefore, besides regular waves and shallow water solitons, the nonlinear solitary groups are a representative wave structure in the sea. The pressure induced by these groups is in focus of this study.

Our approach is not limited by the simplified formulation suggested by Longuet-Higgins (1950). The wave dispersion is taken into account by exact solution of the Laplace equation for potential waves. The surface boundary conditions which correspond to the nonlinearly generated long perturbations are obtained with the help of the weakly nonlinear asymptotic approach for modulated wave trains. The focus on narrow frequency spectra allows us to simplify the approach by Hasselmann (1963) and to obtain eventually closed form expressions for the solution which may be straightforwardly solved numerically with the help of the Fourier transform subroutine. For the case of wave groups having the shape of the envelope solitons of the nonlinear Schrödinger equation, analytic solutions in terms of special functions are obtained.

The paper is organized as follows. The classic framework of potential planar water waves is given briefly in Sec. 2. In particular the solution of the Laplace equation with the help of the Fourier method is outlined. The asymptotic theories for the large-scale perturbations induced by unidirectional and counter propagating modulated waves are given in Sec. 3. They determine the surface boundary conditions for the Laplace equation. The details of the derivation of the theory for opposite waves are collected in Appendix A. The resulting formulas for the underwater dynamic pressure are given in Sec. 3. They require the knowledge of the space-time evolution of the squared wave envelope. The particular case of the wave groups having sech-shapes of the nonlinear Schrödinger equation envelope solitons is discussed in Sec. 4 in details.

## 2. The pressure field in the volume of potential surface waves

We confine the consideration to the framework of the two-dimensional potential Euler equations for ideal incompressible inviscid fluid. Then the motions of the fluid in the water volume from the surface $\eta(x, t)$ to the flat bottom $z = -h$ are governed by the Laplace equation on the velocity potential $\varphi(x, z, t)$,

$$\frac{\partial^2 \varphi}{\partial x^2} + \frac{\partial^2 \varphi}{\partial z^2} = 0, \qquad -h \leq z \leq \eta(x,t). \tag{1}$$

The boundary condition at the bottom requires a zero vertical component of the velocity,

$$\frac{\partial \varphi}{\partial z} = 0 \quad \text{at} \quad z = -h, \tag{2}$$

which in the infinitely deep water limit, $h \to \infty$, transforms to the condition $\varphi \to 0$. The surface boundary condition is nonlinear and consists of the dynamical and kinematic conditions respectively,

$$g\eta + \frac{\partial \varphi}{\partial t} + \frac{1}{2}\left(\left(\frac{\partial \varphi}{\partial x}\right)^2 + \left(\frac{\partial \varphi}{\partial z}\right)^2\right) = 0 \quad \text{at} \quad z = \eta, \tag{3}$$

$$\frac{\partial \varphi}{\partial z} = \frac{\partial \eta}{\partial t} + \frac{\partial \eta}{\partial x}\frac{\partial \varphi}{\partial x} \quad \text{at} \quad z = \eta, \tag{4}$$

where $g$ is the acceleration due to gravity.

If the velocity potential at the rest water level is specified, $\Phi(x, t) \equiv \varphi(x, z = 0, t)$, then the potential in the entire volume occupied by water may be straightforwardly obtained with the help of the Fourier method,

$$\varphi(x,z,t) = \hat{F}^{-1}\left\{\hat{\Phi}(\omega,k)\frac{\cosh(k(z+h))}{\cosh(kh)}\right\} \tag{5}$$

where the double Fourier direct and inverse transforms are defined in the form

$$\hat{F}\{r(x,t)\} \equiv \frac{1}{(2\pi)^2}\iint r(x,t)e^{-i\omega t + ikx}dxdt = \hat{r}(\omega,k),$$

$$\hat{F}^{-1}\{\hat{r}(\omega,k)\} = \iint \hat{r}(\omega,k)e^{i\omega t - ikx}d\omega dk = r(x,t). \tag{6}$$

In (6) the integration is assumed to perform over infinite domains in time and space. One may note that the Laplace equation does not contain time dependence, thus the solution inherits it from the surface velocity potential. Also, the double Fourier transformation may be replaced by the transformation in either time or space if the relation between the wavenumbers and frequencies is given.

The present study is focused on relatively large depth. In the limit of very deep water the vertical structure of the modes may be simplified, and then formula (5) transforms to

$$\varphi(x,z,t) = \hat{F}^{-1}\left\{\hat{\Phi}(\omega,k)e^{|k|z}\right\} \tag{7}$$

When the pressure at the surface is assumed to be zero, the total pressure in the water volume $P_{tot}$ is defined by the Bernoulli law

$$\frac{1}{\rho}P_{tot} = -\frac{\partial\varphi}{\partial t} - \frac{1}{2}\left(\left(\frac{\partial\varphi}{\partial x}\right)^2 + \left(\frac{\partial\varphi}{\partial z}\right)^2\right) - gz, \tag{8}$$

where the constant $\rho$ is the water density ($P_{tot} = 0$ at $z = \eta$ according to (3) The normalized dynamic pressure which characterizes the excess of the total pressure over the hydrostatic pressure is given by

$$p \equiv \frac{1}{\rho}P_{tot} + gz = -\frac{\partial\varphi}{\partial t} - \frac{1}{2}\left(\left(\frac{\partial\varphi}{\partial x}\right)^2 + \left(\frac{\partial\varphi}{\partial z}\right)^2\right). \tag{9}$$

The two last terms in (9) are nonlinear and may be neglected compared to the time derivative of $\varphi$, if the wave amplitude is small. In what follows the dynamic pressure will be approximated by the leading-order part of (9), i.e.

$$p \approx -\frac{\partial\varphi}{\partial t}. \tag{10}$$

Note that the approximation (10) is applied merely for simplicity; the exact dynamic pressure (9) may be easily obtained when the velocity potential $\varphi(x, z, t)$ is known.

## 3. The weakly nonlinear theory for long-scale perturbations induced by modulated waves

To have a reference solution, we start with the classic case of a uniform progressive linear wave. If the surface velocity potential is a harmonic function

$$\Phi_{pw}(x,t) = -\frac{g}{\omega_0}A_{pw}\sin(\omega_0 t - k_0 x), \tag{11}$$

then the potential and the linear pressure in the water bulk are given by (5) and (10) as

$$\varphi_{pw}(x,z,t) = -\frac{g}{\omega_0}A_{pw}\sin(\omega_0 t - k_0 x)\frac{\cosh(k_0(z+h))}{\cosh(k_0 h)}, \tag{12}$$

$$p_{pw}(x,z,t) = gA_{pw}\cos(\omega_0 t - k_0 x)\frac{\cosh(k_0(z+h))}{\cosh(k_0 h)}, \tag{13}$$

According to (3), the linear solution for the surface displacement reads

$$\eta_{pw}(x,t) = A_{pw}\cos(\omega_0 t - k_0 x), \tag{14}$$

and hence the parameter $A_{pw}$ is the amplitude of the plane wave. Here $k_0$ and $\omega_0$ are the wavenumber and the cyclic frequency related according to the linear dispersion relation

$$\omega_0^2 = k_0 g\sigma, \quad \sigma \equiv \tanh k_0 h. \tag{15}$$

In the deep water regime formulas (12), (13) become simpler, the vertical attenuation is described by the exponential factor $\propto \exp(|k_0|z)$, and thus the characteristic vertical scale of the dynamic pressure associated with a uniform progressive wave is of the order of the wave length, $k_0^{-1}$.

The nonlinear wave interactions lead to the virtual emergence of new combinational scales due to the bound harmonics, including induced large-scale motions. The Laplace equation (1) possesses similar scales for horizontal and vertical disturbances, and thus the longer wave components are expected to penetrate deeper into the water.

With the purpose to consider the effect of induced large-scale motions due to the wave interactions we represent the velocity potential and the surface displacement as combinations of two constituents,

$$\varphi = \widetilde{\varphi} + \overline{\varphi}, \qquad \eta = \widetilde{\eta} + \overline{\eta}, \qquad (16)$$

where the terms with tildes denote the primary wave components with characteristic wave lengths $O(1)$. Terms $\widetilde{\eta}$ and $\widetilde{\varphi}$ will be assumed small compared to the inverse dominant wave scale, $k_0^{-1}$, and to the celerity of water particles, correspondingly, as the water waves are often small-amplitude. The weak nonlinearity is characterized by the parameter $\varepsilon \ll 1$, which corresponds to the wave steepness, $k_0 \widetilde{\eta} = O(\varepsilon)$, $k_0^2/\omega_0\, \widetilde{\varphi} = O(\varepsilon)$. The terms with overlines in (16) are the induced components characterized by a large spatial scale along $Ox$ (larger than $k_0^{-1}$).

### 3.1. Progressive nonlinear groups

For weakly nonlinear weakly modulated waves with the same small parameter $\varepsilon$ characterizing the nonlinearity and dispersion, the classic nonlinear Schrödinger theory (NLS) is valid (e.g., Slunyaev, 2005). The leading-order solution reads

$$\widetilde{\varphi}(x,z,t) = i\frac{g}{2\omega_0}(AE - c.c.)\frac{\cosh(k_0(z+h))}{\cosh(k_0 h)} + O(\varepsilon^2), \qquad E = \exp(i\omega_0 t - ik_0 x), \qquad (17)$$

$$\widetilde{\eta}(x,t) = \frac{1}{2}(AE + c.c.) + O(\varepsilon^2), \qquad (18)$$

$$\widetilde{p}_{NLS}(x,z,t) = \frac{g}{2}(AE + c.c.)\frac{\cosh(k_0(z+h))}{\cosh(k_0 h)} + O(\varepsilon^2), \qquad (19)$$

what is similar to (11)-(14) but the complex envelope $A(x, t)$ is a slow function of the horizontal coordinate and time. As before, $k_0$ and $\omega_0$ are related by (15).

The nonlinear wave interaction induces a long-scale current (and the associated surface displacement), which is a slow function of time and space specified by the following relation (Slunyaev, 2005)

$$\frac{\partial \overline{\varphi}}{\partial x} = \gamma |A|^2 + O(\varepsilon^3), \qquad -h \leq z \leq 0, \qquad (20)$$

$$\gamma \equiv \frac{g^2}{\omega_0^2}\frac{k_0^2 C_{gr}(\sigma^2 - 1) - 2\omega_0 k_0}{4(gh - C_{gr}^2)}, \qquad C_{gr} = \frac{g}{2\omega_0}\left(\sigma + k_0 h(1 - \sigma^2)\right).$$

In (20) the value of $\overline{\varphi}$ is of order of $O(\varepsilon)$, though the spatial derivative is one order smaller. The quantity $C_{gr}$ is the group velocity, which tends to $C_{gr} \to (gh)^{1/2}$ in the shallow water limit $k_0 h \ll 1$. Hence for a given depth $h$ the right-hand-side of (20) becomes enormously large in the limit of very long waves and the NLS expansions become invalid.

Note that the solution (20) for the induced current, and, correspondingly, for the dynamic pressure, has no dependence on depth at all, thus when $k_0 h \to -\infty$ the classic NLS theory predicts infinitely deep penetration of the dynamic pressure due to nonlinear wave interactions. In this sense the theory is similar to the approach by Longuet-Higgins (1950). The inability of description of the large-scale induced wave components is a known drawback of the NLS theory, which was remediated in the theory by Dysthe (1979). The Dysthe approach allows describing the mean flow; it is valid for sufficiently deep water, when the simplified shape of the vertical wave mode may be utilized.

The two components of the velocity potential, $\tilde{\varphi}$ and $\overline{\varphi}$ should satisfy the linear Laplace equation independently. In the Dysthe (1979) theory the problem on $\overline{\varphi}$ consists of the Laplace equation below the rest level with the vanishing condition from below

$$\frac{\partial^2 \overline{\varphi}}{\partial x^2} + \frac{\partial^2 \overline{\varphi}}{\partial z^2} = 0, \qquad z \leq 0, \tag{21}$$

$$\overline{\varphi} \to 0, \qquad z \to -\infty, \tag{22}$$

and the nonlinear boundary condition at the rest level which follows from the asymptotic expansions of the surface boundary conditions (3) and (4),

$$\frac{\partial \overline{\varphi}}{\partial z} = \frac{\omega_0}{2} \frac{\partial}{\partial x} |A|^2 + O(\varepsilon^4) \quad \text{at} \quad z = 0. \tag{23}$$

The deep-water dispersion relation follows from (15); it holds

$$\omega_0^2 = k_0 g. \tag{24}$$

According to the Dysthe theory, values of the induced velocity potential $\overline{\varphi}$ are of the order $O(\varepsilon^2)$, and the corresponding velocities (horizontal and vertical) are one order smaller due to the large horizontal and associated vertical scales.

Thus, the vertical derivative of the potential at the rest level is specified by (23), and the problem (21)-(23) may be easily solved with the help of the Fourier method (7) with use of the relation

$$\hat{F}\left\{ \left. \frac{\partial \varphi}{\partial z} \right|_{z=0} \right\} = |k|\hat{\Phi}(\omega, k), \tag{25}$$

where $\hat{\Phi}$ is the Fourier transform of the velocity potential at the water rest level.

In the leading order the derivation by time may be changed as $\partial/\partial t \to -C_{gr}\partial/\partial x$, and then (10) results in a trivial relation between the induced pressure and the induced mean current $\overline{u}$ which depends on depth (see also discussion of this problem in McIntyre, 1981),

$$\overline{p} \approx C_{gr}\overline{u}, \qquad \overline{u} = \frac{\partial \overline{\varphi}}{\partial x}. \tag{26}$$

Then with use of equations (7) and (23) the main contribution to the induced large-scale dynamic pressure is given by the following expression,

$$\overline{p}_{unidir} = -\frac{\omega_0^2}{4k_0}\hat{F}^{-1}\left\{ |k| f e^{|k|z} \right\}, \quad z \leq 0, \qquad f(\omega, k) \equiv \hat{F}\left\{ |A|^2 \right\}, \tag{27}$$

which imply sufficiently deep water. Its magnitude is of the order $O(\varepsilon^3)$. The general solution (27) may be further simplified considering the initial or boundary-value problem, see equation (36) below.

### 3.2 Counter propagating nonlinear wave groups

Longuet-Higgins (1950) noticed that two uniform waves propagating in the opposite directions generate a second-order component which does not decay with depth and oscillates with the double frequency (within the potential hydrodynamic theory). The corresponding induced pressure oscillations are of the order $O(\varepsilon^2)$, what is one order larger than due to the nonlinear wave modulations (27) discussed above.

In this section we consider two wave systems with a narrow frequency spectrum which propagate in the opposite directions, and thus generalize the solution by Longuet-Higgins (1950). We apply a similar approach as Dysthe (1979) based on the asymptotic expansions of weakly nonlinear weakly modulated waves in water with a constant depth. The derivation is performed in the manner similar to (Slunyaev, 2005) and is given in Appendix A.

The counter propagating waves are represented by complex envelopes $A_+(x, t)$ and $A_-(x, t)$ for the surface displacement $\tilde{\eta}$ and potential $\tilde{\varphi}$ (similar to the unidirectional case (17)-(18)), where the low indices denote the direction of propagation (along $Ox$ or opposite to it). The trains are characterized by similar wavenumbers $k_0$ and frequencies $\omega_0$ and $-\omega_0$ related by the dispersion law (15). The induced velocity potential $\overline{\varphi}$ at $z = 0$ satisfies the Laplace equation (21) with a nonleaking condition at the bottom, and is determined by the upper boundary condition (see (A.10) in Appendix A)

$$g\frac{\partial \overline{\varphi}}{\partial z} + \frac{\partial^2 \overline{\varphi}}{\partial t^2} = i\frac{1+3\sigma^2}{2}\omega_0^3\left[A_+ A_-^* \exp(2i\omega_0 t) - c.c.\right] \quad \text{at} \quad z = 0. \quad (28)$$

It is a slow function of the depth. Note that the Dysthe theory for unidirectional waves requires consideration of the next order asymptotic expansion than (28).

In the particular case of two uniform waves which travel in opposite directions, amplitudes $A_+$ and $A_-$ are just numbers (i.e., wave amplitudes) and the product $A_+ A_-^*$ is real, when an appropriate time reference is chosen, then (28) yields

$$\overline{\varphi} = \frac{1+3\sigma^2}{4}\omega_0 A_+ A_-^* \sin(2\omega_0 t), \quad -h \leq z \leq 0. \quad (29)$$

This result agrees with the solution by Dalzell (1999) for finite-depth second-order wave–wave interactions. For the upper condition (29) the Laplace equation admits a solution which is uniform with depth. Then the dynamic pressure (10) reads

$$\overline{p} = -\frac{1+3\sigma^2}{2}\omega_0 A_+ A_-^* \cos(2\omega_0 t), \quad -h \leq z \leq 0, \quad (30)$$

it oscillates with the double dominant frequency. Solution (30) in the infinitely deep water case (i.e., when $\sigma = 1$) coincides with the formula by Longuet-Higgins (1950) (different orientation of the $Oz$ axis in (Longuet-Higgins, 1950) and in the present formulation should be taken into account though).

Again, the Fourier method (7) helps to write down the solution for the velocity potential in the water volume with the condition at $z = 0$ in form (28). For the sake of simplicity we give here the solution for the dynamic pressure in the deep-water limit, $\sigma = 1$,

$$\overline{p}_{opposite} = -2\omega_0^3 \hat{F}^{-1}\left\{\omega\frac{f(\omega - 2\omega_0, k) - f^*(-\omega - 2\omega_0, -k)}{\omega^2 - g|k|}e^{|k|z}\right\}, \quad z \leq 0, \quad (31)$$

$$f(\omega, k) = \hat{F}\{A_+ A_-^*\}.$$

The classic result of Longuet-Higgins (1950) for a uniform wave (30) follows when the second term in the denominator in (31), $g|k|$, is neglected. Thus the full solution for weakly modulated patterns (31) is expected to generally exceed the estimate (30) at the surface. This remark holds for the velocity potential field as well (see discussion of Fig. 4 in the next section).

**4. Deep-penetrating pressure fields caused by sech-shape wave groups**

A few examples of waves and induced dynamic pressure fields are considered in this section. The infinitely deep water limit is considered for simplicity. A standing wave was concerned by Longuet-Higgins (1950), which we use as a classic reference case. Solitary groups are another important case of sea wave trains supported by the balance between the nonlinear and dispersive effects, which are convenient for the study. Envelope solitons are exact solutions of the nonlinear Schrödinger equation (Zakharov, 1968), which is valid for long modulations of small amplitude waves. At the same time this approximate solution describes reasonably well even short solitary groups of steep planar waves (Slunyaev et al., 2013, 2017). Thus, an exact envelope soliton of the NLS equation is the first approximation to typical sea groups.

The analytic solution of the nonlinear Schrödinger equation in terms of the complex envelope reads

$$A = \frac{a}{\cosh\left[\sqrt{2}ak_0^2(x - C_{gr}t)\right]} \exp\left[i\frac{(k_0 a)^2}{4}\omega_0 t\right]. \tag{32}$$

It represents a wave group with amplitude $a$, dominant wavenumber $k_0$ and frequency $\omega_0$ corrected by the nonlinear shift (due to the exponential term in (32)), which propagates in the direction of $Ox$ with the velocity of linear waves $C_{gr} = \omega_0/k_0/2$. Within the NLS framework the envelope shape is preserved if the group propagates isolated, and restores after collisions with all other waves. For the sake of simplicity the nonlinear frequency shift will be neglected, and groups with the following momentary shape will be considered below,

$$A_{sol} = \frac{a}{\cosh\left[\sqrt{2}ak_0^2 x\right]}. \tag{33}$$

The spatial Fourier transforms of $A_{sol}$ (33) and $A_{sol}^2$ may be calculated analytically. Two distinct situations are considered below:

1) a solitary group (33) which propagates with the constant velocity $C_{gr}$;
2) a collision of two counter propagating groups, each having the shape (33) with the same amplitudes, $A_+ = A_- = A_{sol}$.

**Analytic solution**

The induced pressure in the field of unidirectional waves is given by formula (27), which uses the Fourier transform of the squared envelope $|A|^2$ in the space-time domain. If the group propagates with the constant velocity $C_{gr}$ as a whole, the Fourier transform *for the envelope* may be represented in the form

$$f(\omega, k) = f_0(k)\delta(\omega - kC_{gr}), \tag{34}$$

which is convenient for consideration of the Cauchy problem. In (34) $\delta(\cdot)$ denotes the Dirac delta function. Function $f_0(k)$ is the spatial Fourier transform of the squared envelope $|A|^2$ at the moment $t = 0$,

$$f_0 = \frac{1}{2\pi}\int |A(t=0)|^2 e^{ikx} dx, \tag{35}$$

consistently with (6). After the integration of (27) over frequencies, one obtains the general solution for the Cauchy problem for unidirectional stationary groups

$$\overline{p}_{unidir} = -\frac{\omega_0^2}{4k_0}\int_{-\infty}^{\infty} |k| f_0 \exp(ikC_{gr}t - ikx) e^{|k|z} dk, \tag{36}$$

which may be further reformulated in terms of the spatial inverse Fourier transform. In fact, solution (36) is valid in the general case of weakly modulated weakly nonlinear waves considered in Sec. 3.1, when relation (34) is approximate.

The spatial Fourier transform (5) for the group (33) possesses an analytic representation thanks to the table of integrals by Gradshteyn & Ryzhik (2007; item 3.982)

$$f_0 = \frac{1}{4k_0^4}\frac{k}{\sinh\dfrac{\pi k}{2\sqrt{2}k_0^2 a}}. \tag{37}$$

Then, in a few manipulations solution (36) for a single solitary group (33) results in

$$\bar{p}_{one} = -\frac{g}{8k_0^4} \int_0^\infty \frac{k^2 \cos k(x - C_{gr}t)}{\sinh \frac{\pi k}{2\sqrt{2}k_0^2 a}} e^{kz} dk . \tag{38}$$

Solution (38) represents the induced pressure field which co-propagates with the group and does not vary in the co-moving reference system.

In the case of collision of two counter propagating wave groups the solution (31) should be used. Assuming that the spectrum of the group $f(\omega, k)$ is narrow in both frequency and wavenumber domains, solution (31) may be reduced to

$$\bar{p}_{opposite} \approx -2\omega_0^2 \cos 2\omega_0 t \int f_0 e^{|k|z} e^{-ikx} dk , \tag{39}$$

(see details in Appendix B), which is the solution for two colliding wave trains of a given narrow spectrum (35) for $A_+$ and $A_-$. If each of them have the shape (33), the spectrum function $f_0$ at the moment of collision $t = 0$ is specified by (37), and then (39) yields

$$\bar{p}_{two} = -\frac{g}{k_0^3} \cos(2\omega_0 t) \int_0^\infty \frac{k \cos kx}{\sinh \frac{\pi k}{2\sqrt{2}k_0^2 a}} e^{kz} dk . \tag{40}$$

It may be seen from (38) and (40) that the characteristic horizontal and vertical scales of the solutions correspond to the length of the group (33). Solutions (38) and (40) allow the representation in terms of the polygamma functions (Gradshteyn & Ryzhik, 2007) $\psi^{(m)}(\xi) = d^{m+1}/d\xi^{m+1} \ln\Gamma(\xi)$, $\psi^{(0)} \equiv \psi = \ln\Gamma$, where $\Gamma(\xi)$ is the gamma function, see equalities (B.5) and (B.6) in Appendix B. At location $x = 0$ the solutions read

$$\bar{p}_{one}(x=0) = \frac{1}{\sqrt{2}\pi^3} \frac{g}{k_0} (k_0 a)^3 \psi''\left(\frac{1}{2} - \frac{\sqrt{2}}{\pi} ak_0^2 z\right), \tag{41}$$

$$\bar{p}_{two}(x=0) = -\frac{4}{\pi^2} \frac{g}{k_0} (k_0 a)^2 \psi'\left(\frac{1}{2} - \frac{\sqrt{2}}{\pi} ak_0^2 z\right). \tag{42}$$

The difference in magnitude of the dynamic pressure under one progressive and two colliding solitary groups may be estimated as follows

$$\left.\frac{\bar{p}_{two}}{\bar{p}_{one}}\right|_{x=0} = -\frac{4\sqrt{2}\pi}{k_0 a} \frac{\psi'\left(\frac{1}{2} - \frac{\sqrt{2}}{\pi} ak_0^2 z\right)}{\psi''\left(\frac{1}{2} - \frac{\sqrt{2}}{\pi} ak_0^2 z\right)}, \tag{43}$$

see the solid line in Fig. 1a. At the water rest level, $z = 0$, (43) gives approximately $5.21(k_0 a)^{-1}$, and the coefficient grows with depth almost linearly. The dashed line in Fig. 1a corresponds to the relation

$$\left.k_0 a \frac{\bar{p}_{two}}{\bar{p}_{one}}\right|_{x=0} \cong -4\sqrt{2}\pi\left(\frac{\psi'(0.5)}{\psi''(0.5)} + \frac{\sqrt{2}}{\pi} ak_0^2 z\right) \approx 5.21 - 8ak_0^2 z . \tag{44}$$

Which approximates (43) fairly well. It has been mentioned before that steeper envelope solitons are shorter, and thus the characteristic scale of the wave fields' attenuation with $|z|$ is also shorter. However, according to solutions (41) and (42), the peak dynamic pressure under a single or colliding groups at a given depth $z$ increases for steeper waves, see Fig. 1b,c.

## Numerical calculation of the pressure fields

Below we illustrate few cases, where the solutions for the induced large-scale pressure are obtained by virtue of the numerical integration of solutions (27), (31) with the help of the fast Fourier transform in time and space. Thus, the simplifications implied in the solutions

(39) and (40) are not employed. The situations of a uniform wave, one solitary wave group (33), and two head-on colliding groups (33) are considered in this section. Two wave steepnesses are considered in the examples, $k_0 a = 0.1$ and $k_0 a = 0.25$.

The surface displacements which in the first approximation (just the linear superposition) correspond to the two colliding groups (33) with the similar steepness are shown in Fig. 2 at different phases: $\omega_0 t = 0$ (zero displacement), $\omega_0 t = \pi/4$ and $\omega_0 t = \pi/2$ (maximum displacement). For the convenience, the phases in Fig. 2 are given in terms of $\cos 2\omega_0 t$. Single unidirectional groups of a given steepness have twice smaller amplitudes than shown in Fig. 2. The steeper group is significantly shorter, providing the approximate balance of the dispersive and nonlinear effects.

The dependencies of the dynamic pressure with depth at $x = 0$ are shown in Fig. 3 for a progressive wave with amplitude $a$, one solitary group and two colliding groups with amplitudes $a$. The instants when the pressure is maximal are considered. Two cases of steepness are shown in Fig. 3a,b. Note that the curves for single groups represent the values multiplied by the factor 55 in Fig. 3a and by the factor 23 in Fig. 3b with the purpose to have at $z = 0$ the same magnitudes as for the colliding groups (see (43)); these values define the upper limits for the horizontal axes in Fig. 3. The pressure of progressive waves is characterized by an exponential profile (red curves with dots in Fig. 3) with maxima at $z = 0$ equal to $ga$ (see (13)), which exceed the values of the induced pressure beneath the groups. At the same time the induced pressure for the steep colliding groups is only about two times smaller than the maximum pressure for the linear uniform wave (Fig. 3b). The dynamic pressure beneath the single group decays much slower than the pressure caused by a uniform wave. The pressure field beneath the colliding groups is noticeably stronger and decays even slower than in the case of a single group.

Despite the fact that the pressure induced by colliding groups is much larger than that of a single group, the velocity potential profiles at the rest level $z = 0$ are not so much different in magnitudes, see Fig. 4. The induced current is independent on the wave phase in the case of the single solitary group (blue solid curve). The large-scale induced velocity potential profiles for colliding groups oscillate with the frequency $2\omega_0$, and thus three phases are shown in Fig. 4 (thin black lines, see the legend coding; the induced velocity is zero when $\sin 2\omega_0 t = 0$). The reference case of standing waves (30) is also shown (red curves; zero when $\sin 2\omega_0 t = 0$). As has been mentioned before, the maximum of the velocity potential at $z = 0$ in the case of two colliding groups is slightly larger than in the case of two counter propagating uniform waves of the same steepness (better seen in Fig. 4b).

The momentary pressure distributions in the water bulk at the moments of peak values are shown in Fig. 5 (single solitary groups) and Fig. 6 (colliding groups). One may see that the induced pressure is larger in magnitude in steeper cases (Fig. 5b, 6b), note different color bar limits. The biggest values are attained in the situation of opposite wave groups (Fig. 6) in agreement with (43). As smaller amplitude groups are broader, their pressure fields decay with depth slower. However, due to the nonlinear nature of the induced large-scale pressure, as mentioned earlier (Fig. 1b,c), for a given depth the pressure grows as the steepness increases. The pressure imprints of unidirectional and opposite groups are qualitatively different, what may be explained by different profiles of the velocity potential at the surface (Fig. 4). While the dynamic pressure induced by colliding groups is characterized by one polarity in the entire water volume, the pressure beneath unidirectional groups is sign-changing (see limits of the color bars in Fig. 5, 6).

It is interesting to note that the dynamic pressure does not necessarily decay in magnitude with depth, what may be explained by strong inhomogeneity of the waves along $Ox$. The sign of the vertical gradient of the absolute value of the dynamic pressure, $\text{sign}(\partial/\partial z\, p^2)$ is shown in Fig. 7 for one solitary group, and for two colliding groups. The black areas correspond to the negative gradient, i.e., when the pressure at deeper layers is even larger than above. The areas are qualitatively different for the situations of unidirectional and colliding groups.

The described peculiarities of pressure distributions make possible nontrivial patterns of pressure time series measured at different depths. Vertical distributions of the dynamic pressure as functions of time at different locations $x$ from the groups are displayed in Fig. 8 for the situation of two colliding trains. One may see that at different distances from the groups the maximum of pressure variations may be near the surface (Fig. 8a), at large depth (Fig. 8c) or at the intermediate depth (Fig. 8b). The time evolution of the pressure beneath one solitary group retrieved at a single point also looks complicated, see Fig. 9. The depth where the pressure maximum occurs may change with time differently, depending on the horizontal distance to the group.

## 5. Conclusion

In this paper we develop the theory for deep-penetrating into the water column pressure fields beneath intense surface waves. The theory is limited by potential fluid motions and uses the exact solution of the Laplace equation with the help of the Fourier transformation. The induced long-scale perturbations at the water surface due to the nonlinear interactions of the modulated waves are calculated within the asymptotic nonlinear theories for collinear waves. The Dysthe (1979) model is used for unidirectional waves, and a similar theory is developed for counter propagating weakly modulated waves with the same dominant frequencies.

The underwater dynamic pressure fields are discussed in detail for the representative cases of single and head-on colliding groups with sech-shapes typical for the envelope solitons of the nonlinear Schrödinger equation. It is shown that the induced long-scale pressure fields can penetrate deep into the water column for tens of dominant wave lengths for the typical sea wave conditions (steepness $ka = O(10^{-1})$). Its magnitude at the surface may be comparable with the pressure in linear uniform waves, when groups collide and their steepness is large. Thus, intense wave groups may cause unexpectedly extreme impact at deep water layers, especially when interact. Though intense solitary trains get shorter, they produce even stronger dynamic pressure at a given depth. The collisions of solitary groups may be expected to occur in the open sea under the conditions favorable for appearance of the wave modulations due to the Benjamin – Feir instability, or in the coastal zone when the groups may reflect from cliffs.

The patterns of the underwater pressure are complicated; they possess qualitatively different features for the situations of single and colliding wave groups. Aside from the groups the dynamic pressure at deep layers may be larger than above, what may make the problem of reconstruction of the surface movement from the underwater pressure time series even more complicated.

**Acknowledgements**
The research for AS and EP was supported by the Russian Foundation for Basic Research (bilateral grant No. 16-55-52019), the President grant for leading scientific schools SC-6637.2016.5, and also by the Volkswagen Foundation. HH acknowledges the grant support from 105-2923-E-006-002- MY3.

**Appendix A. Theory for two wave systems propagating in opposite directions in water of constant depth**

The hydrodynamic fields are represented in the form of a superposition of the wave part and the induced large-scale part, see (16). Adapting the notations in Slunyaev (2005), the leading order wave field is assumed to have the form

$$\tilde{\varphi}(x,z,t) = \varepsilon \frac{1}{2}\left(B_+(x_1,t_1,z_1)e^{i\omega_0 t_0 - ik_0 x_0} + B_-(x_1,t_1,z_1)e^{-i\omega_0 t_0 - ik_0 x_0} + c.c.\right)\frac{\cosh(k_0(z_0+h))}{\cosh(k_0 h)}, \quad (A.1)$$

$$\tilde{\eta}(x,t) = \varepsilon \frac{1}{2}\left(A_+(x_1,t_1)e^{i\omega_0 t_0 - ik_0 x_0} + A_-(x_1,t_1)e^{-i\omega_0 t_0 - ik_0 x_0} + c.c.\right). \quad (A.2)$$

The fluid motion is represented by two wave systems characterized by complex envelopes for the potential, $B$, and the displacement, $A$, which propagate with the same dominant wavenumber, $k_0$, and frequency, $\omega_0$, in opposite directions indicated by the subscripts $\pm$. The hierarchy of multi-scale variables is introduced with the use of the small parameter $\varepsilon$ as

$$\frac{\partial}{\partial t} = \frac{\partial}{\partial t_0} + \varepsilon \frac{\partial}{\partial t_1} + \ldots, \quad \frac{\partial}{\partial x} = \frac{\partial}{\partial x_0} + \varepsilon \frac{\partial}{\partial x_1} + \ldots, \quad \frac{\partial}{\partial z} = \frac{\partial}{\partial z_0} + \varepsilon \frac{\partial}{\partial z_1} + \ldots \quad (A.3)$$

Thus, the zero subscripts denote the fast variables. The group structure of the solution is represented by the dependence on slow variables $x_1$, $t_1$; it results in slow variations with depth, which are taken into account by virtue of the slow vertical coordinate $z_1$. The induced components are sought in the following form: $\bar{\eta} = \bar{\eta}(x_1, t_0, t_1)$ and $\bar{\varphi} = \bar{\varphi}(x_1, t_0, t_1, z_1)$.

The wave part of the solution, $\tilde{\varphi}$ (A.1), satisfies the Laplace equation (1) in the leading order. The Laplace equation for the large-scale velocity potential $\bar{\varphi}$ (21) is satisfied with the non-leaking boundary condition at $z = -h$, and the surface boundary condition, which follows from (3) and (4), when the velocity potential is decomposed around $z = 0$ into the Taylor series (see in Slunyaev, 2005). For terms with the fast dependence along $Ox$ the following relations between the wave components are obtained from (3) and (4) in the leading order $O(\varepsilon)$:

$$(gA_+ + i\omega_0 B_+)e^{i\omega_0 t_0} + (gA_- - i\omega_0 B_-)e^{-i\omega_0 t_0} = 0 \quad \text{at} \quad z = 0, \quad (A.4)$$

$$(i\omega_0 A_+ - k_0 \sigma B_+)e^{i\omega_0 t_0} + (-i\omega_0 A_- - k_0 \sigma B_-)e^{-i\omega_0 t_0} = 0 \quad \text{at} \quad z = 0. \quad (A.5)$$

The compatibility condition for (A.4) and (A.5) results in the classic dispersion law and relates the envelopes for the velocity potential and the surface displacement,

$$\omega_0^2 = gk_0\sigma, \quad (A.6)$$

$$B_\pm = \pm i \frac{g}{\omega_0} A_\pm. \quad (A.7)$$

Conditions (3), (4) in order $O(\varepsilon^2)$ yield the equations for the large-scale motions,

$$g\bar{\eta} + \frac{\partial \bar{\varphi}}{\partial t} = \frac{1+3\sigma^2}{4}\omega_0^2\left[A_+ A_-^* \exp(2i\omega_0 t) + c.c.\right] - \frac{1-\sigma^2}{4}\omega_0^2\left(|A_+|^2 + |A_-|^2\right) \quad \text{at} \quad z = 0, \quad (A.8)$$

$$\frac{\partial}{\partial t}\bar{\eta} - \frac{\partial}{\partial z}\bar{\varphi} = 0 \quad \text{at} \quad z = 0. \quad (A.9)$$

Conditions (A.8), (A.9) may be combined when the mean surface displacement is expressed from the first equation and plugged into the second one. Then the following equation on the velocity potential specifies the boundary condition at the water rest level,

$$g\frac{\partial \bar{\varphi}}{\partial z} + \frac{\partial^2 \bar{\varphi}}{\partial t^2} = i\frac{1+3\sigma^2}{2}\omega_0^3\left[A_+ A_-^* \exp(2i\omega_0 t) - c.c.\right] \quad \text{at} \quad z = 0. \quad (A.10)$$

Equation (A.10) determines the induced large-scale velocity potential in terms of the envelope amplitudes, $A_+$ and $A_-$, at the upper boundary, $z = 0$. The potential is subject to the Laplace equation (21) with the non-leaking condition at $z = -h$. Formula (A.8) specifies the corresponding large-scale surface displacement. Note that the Dysthe theory (23) requires consideration of the next order asymptotic expansions.

## Appendix B. Derivation of the explicit solutions for the dynamic pressure.
**Collision of two groups.**

Let us assume that the Fourier transform $f(\omega, k)$ of the envelope is narrow in frequency. Then integral (31) may be approximated by

$$\bar{p} \approx -2\omega_0^3 \iint \frac{2\omega_0}{(2\omega_0)^2 - g|k|} \left[ f(\omega - 2\omega_0, k)e^{2i\omega_0 t} + f^*(-\omega - 2\omega_0, -k)e^{-2i\omega_0 t} \right] e^{|k|z} e^{-ikx} d\omega dk . \quad (B.1)$$

If the wavenumber spectrum of the envelope is assumed to be narrow as well, then values of $g|k|$ may be neglected compared to $2\omega_0$, and (B.1) yields

$$\bar{p} \approx -\omega_0^2 \iint \left[ f(\omega - 2\omega_0, k)e^{2i\omega_0 t} + f^*(-\omega - 2\omega_0, -k)e^{-2i\omega_0 t} \right] e^{|k|z} e^{-ikx} d\omega dk . \quad (B.2)$$

For the real-valued envelope $f^*(-\omega, -k) = f(\omega, k)$; the two terms in (B.2) may be integrated over the frequency, and then (B.2) reduces to

$$\bar{p} \approx -2\omega_0^2 \cos(2\omega_0 t) \iint f(\omega, k) e^{|k|z} e^{-ikx} d\omega dk . \quad (B.3)$$

The integrand in (B.3) does not depend on time; then with the use of (6) solution (B.3) yields the solution of the Cauchy problem (39) for any given spectrum $f_0$ specified by (35).

**Representation of the solutions in terms of polygamma functions.**

The following relations may be obtained with the help of tabulated integrals (Gradshteyn & Ryzhik, 2007; see 4.131.3, 8.36), which express integrals (38) and (40) in terms of polygamma functions:

$$\int_0^\infty \frac{k \cos kx}{\sinh qk} e^{kz} dk = \frac{1}{4q^2} \left[ \psi'\left(\frac{1}{2} + \frac{-z + ix}{2q}\right) + \psi'\left(\frac{1}{2} + \frac{-z - ix}{2q}\right) \right], \quad (B.5)$$

$$\int_0^\infty \frac{k^2 \cos kx}{\sinh qk} e^{kz} dk = -\frac{1}{8q^3} \left[ \psi''\left(\frac{1}{2} + \frac{-z + ix}{2q}\right) + \psi''\left(\frac{1}{2} + \frac{-z - ix}{2q}\right) \right], \quad (B.6)$$

where $q = \dfrac{\pi}{2\sqrt{2}k_0^2 a}$.

Though solutions (B.5), (B.6) are written in the complex form, they represent real-valued solutions. Functions $\psi^{(m)}(z) = d^{m+1}/dz^{m+1} \ln\Gamma(z)$ are $m$-th polygamma functions, $\psi^{(0)} \equiv \psi = \ln\Gamma$, which may be represented in the integral form

$$\psi^{(m)}(z) = (-1)^{m+1} \int_0^\infty \frac{k^m e^{-zk}}{1 - e^{-k}} dk . \quad (B.7)$$

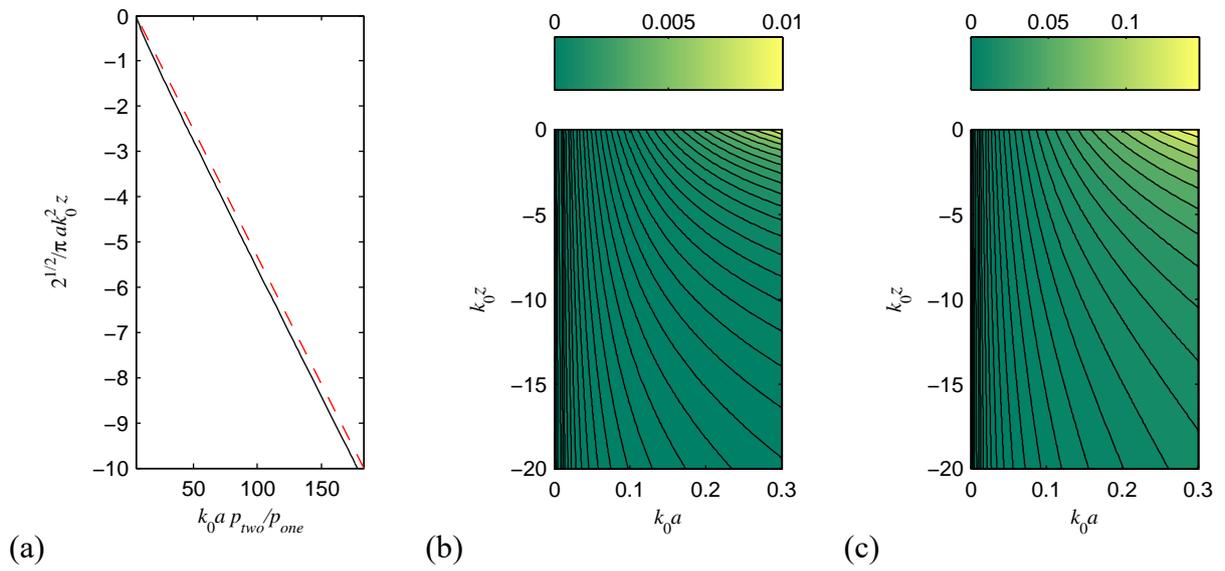

Fig. 1. Pressure under the groups ($t = 0$, $x = 0$) according to the analytic solutions (40) and (41): the ratio of pressure magnitudes under one and two colliding groups (42) (solid line) and approximation (43) (dashed line) (a); and the dependences of pressure $k_0 \bar{p}/g$ on the depth and the wave steepness for one (b) and two (c) groups.

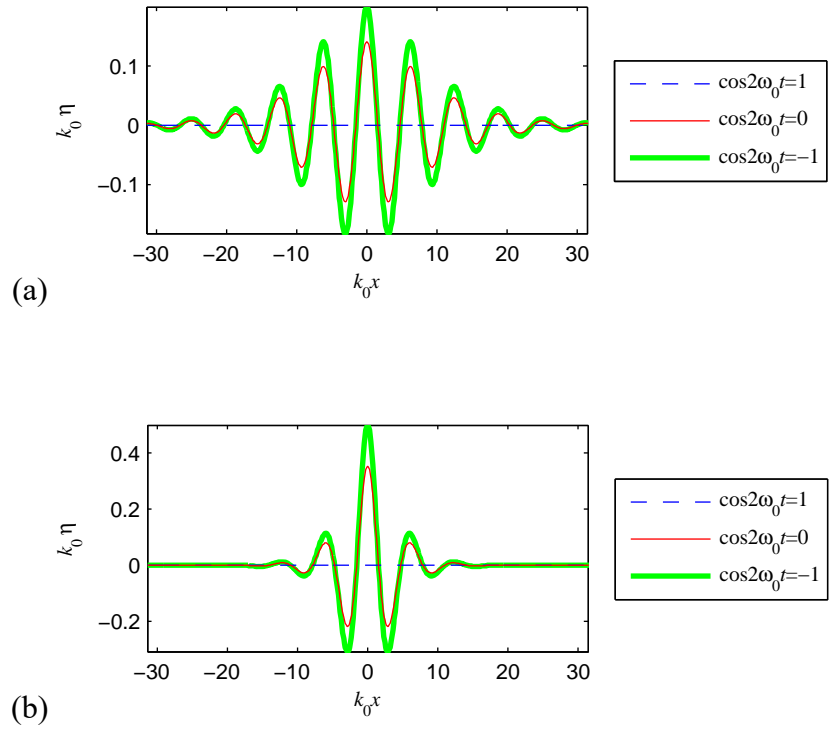

Fig. 2. Surface displacements at the moment of collision of two solitary groups with the steepness $k_0 a = 0.1$ (a) and $k_0 a = 0.25$ (b) at different phases (see the legend).

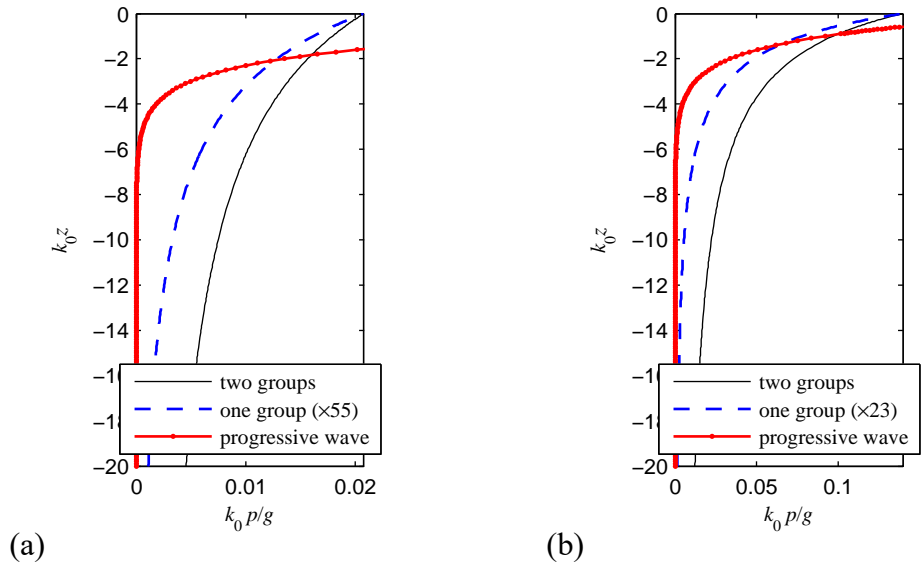

Fig. 3. Dynamic pressure profiles as functions of depth for two wave steepness, $k_0 a = 0.1$ (a) and $k_0 a = 0.25$ (b): uniform progressive waves, one solitary group and two colliding groups (see the legend). Note that the pressures induced by the single groups are amplified by factors 55 (a) and 23 (b) respectively.

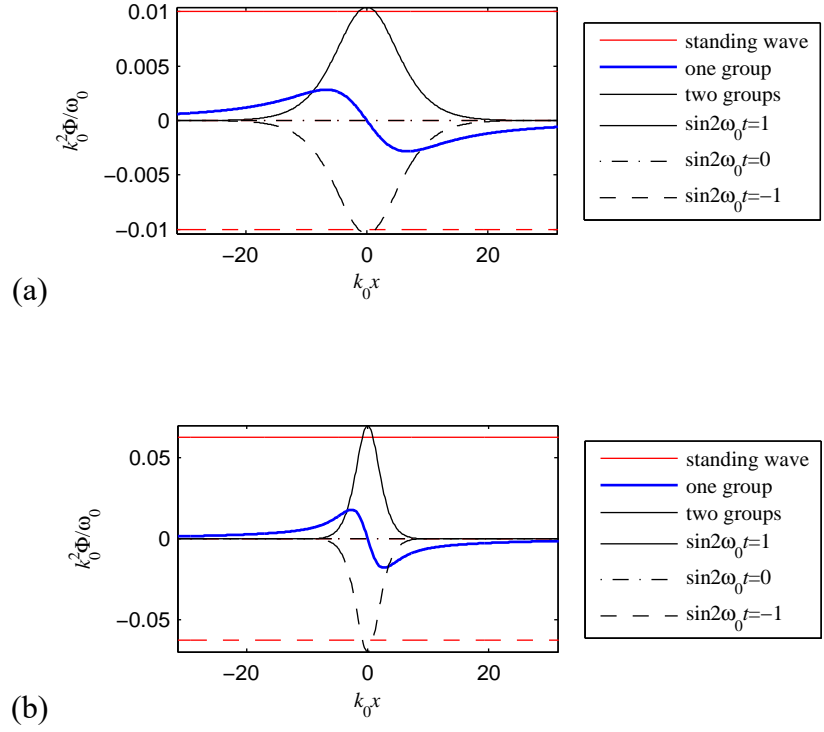

Fig. 4. Velocity potential profiles $\overline{\Phi}(x)$ at $z = 0$ for a single solitary group (blue curve), colliding groups (black curves, three phases are shown, see the legend) and a standing wave (red constant lines, three phases are shown by different lines similar to the case of colliding groups). Two wave steepnesses are considered: $k_0 a = 0.1$ (a) and $k_0 a = 0.25$ (b).

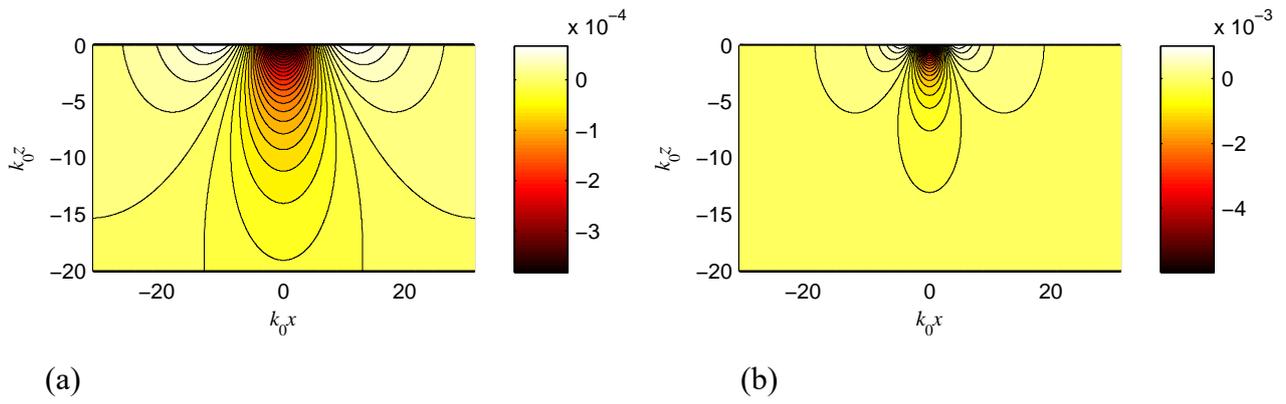

Fig. 5. Momentary large-scale dynamic pressure distributions in the water volume induced by single wave groups of two steepness, $k_0 a = 0.1$ (a) and $k_0 a = 0.25$ (b). Note different maximal magnitudes of the color scales.

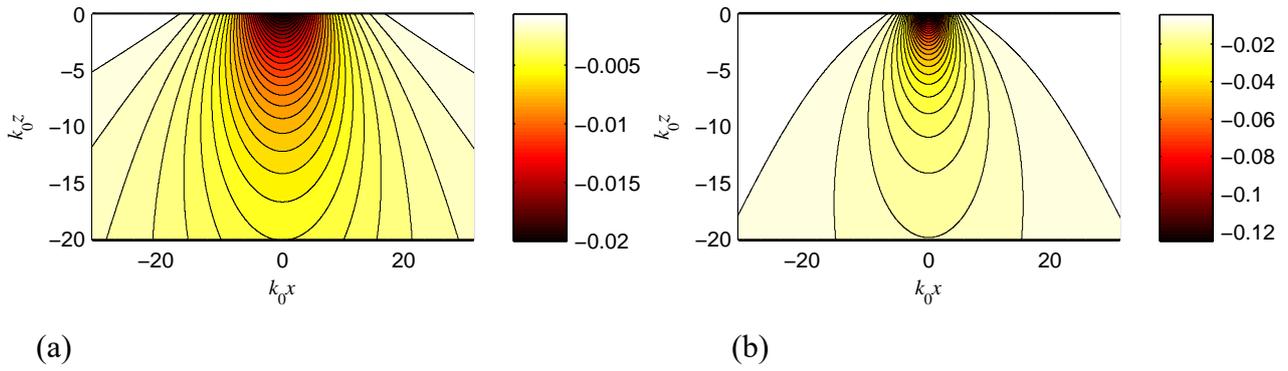

Fig. 6. Similar to Fig. 5, but the case of two colliding wave groups with the steepness $k_0 a = 0.1$ (a) and $k_0 a = 0.25$ (b). The solution at phase $\cos 2\omega_0 t = 1$ is shown.

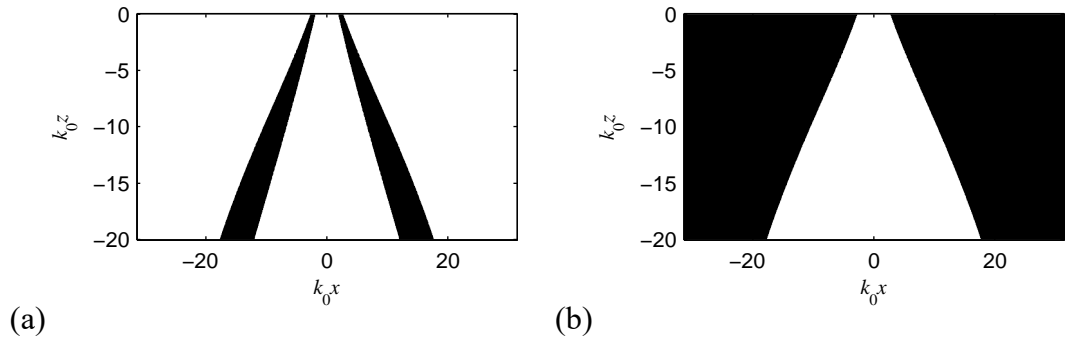

Fig. 7. Sign of the vertical gradient of the absolute value of the induced dynamic pressure for one solitary group (a), and colliding groups (b). Black areas correspond to the negative gradient situation, when the dynamic pressure is larger at deeper layers. Case $k_0 a = 0.1$ is shown.

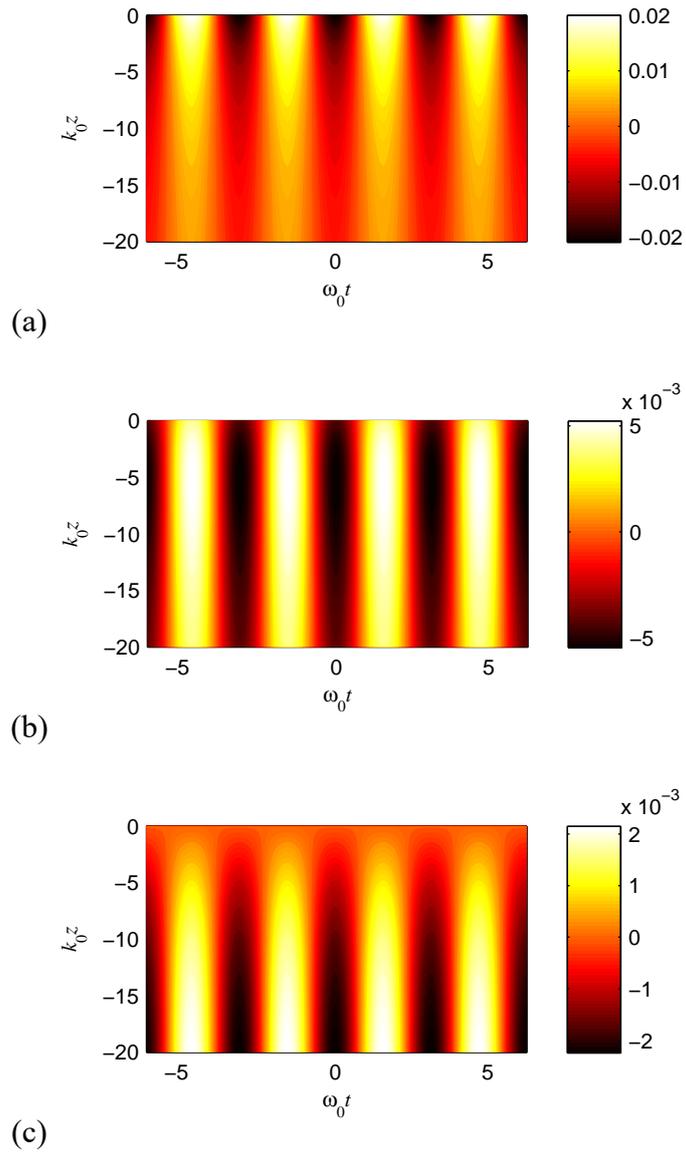

Fig. 8. Dynamic pressure distributions as functions of time recorded at different locations: $x = 0$ (a), $k_0 x = 10$ (b) and $k_0 x = 25$ (c). Two colliding groups with the steepness $k_0 a = 0.1$.

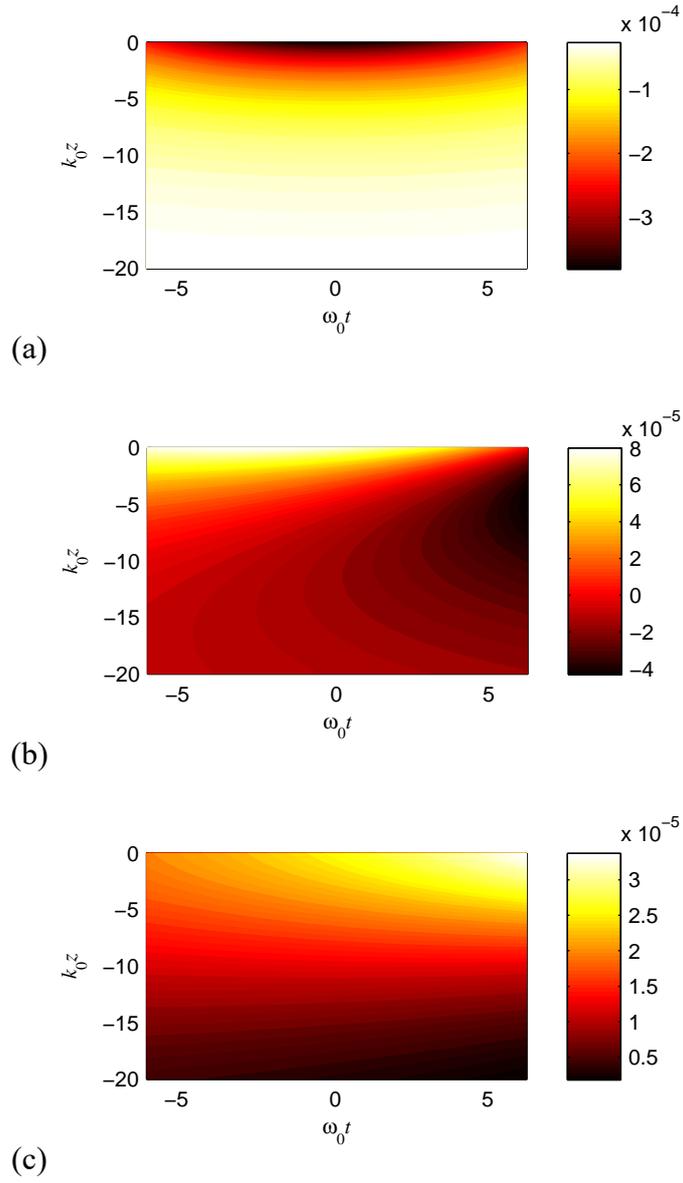

Fig. 9. Dynamic pressure distributions as functions of time recorded at different locations: $x = 0$ (a), $k_0 x = 10$ (b) and $k_0 x = 25$ (c). The case of one solitary group with the steepness $k_0 a = 0.1$ is shown.